# Controlling Excitons in an Atomically Thin Membrane with a Mirror

You Zhou[1,2], Giovanni Scuri[2], Jiho Sung[1,2], Ryan J. Gelly[2], Dominik S. Wild[2], Kristiaan De Greve[1,2], Andrew Y. Joe[2], Takashi Taniguchi[3], Kenji Watanabe[3], Philip Kim[2], Mikhail D. Lukin[2] & Hongkun Park[1,2]*

[1]Department of Chemistry and Chemical Biology, Harvard University, Cambridge, MA 02138, USA

[2]Department of Physics, Harvard University, Cambridge, MA 02138, USA

[3]National Institute for Materials Science, 1-1 Namiki, Tsukuba 305-0044, Japan

*To whom correspondence should be addressed: hongkun_park@harvard.edu

**We demonstrate a new approach for dynamically manipulating the optical response of an atomically thin semiconductor, a monolayer of $MoSe_2$, by suspending it over a metallic mirror. First, we show that suspended van der Waals heterostructures incorporating a $MoSe_2$ monolayer host spatially homogeneous, lifetime-broadened excitons. Then, we interface this nearly ideal excitonic system with a metallic mirror and demonstrate control over the exciton-photon coupling. Specifically, by electromechanically changing the distance between the heterostructure and the mirror, thereby changing the local photonic density of states in a controllable and reversible fashion, we show that both the absorption and emission properties of the excitons can be dynamically modulated. This electromechanical control over exciton dynamics in a mechanically flexible, atomically thin semiconductor opens up new avenues in cavity quantum optomechanics, nonlinear quantum optics, and topological photonics.**

The optical properties of resonant quantum emitters and materials can be controlled via judicious engineering of their local electromagnetic environment [1]. In solid-state devices, such control is often achieved by coupling emitters to optical cavities [2-7]. The coupling between quantum-well excitons and photons in microcavities results in the formation of exciton polaritons [7-15], and the strong coupling of quantum dots to tightly confined photons enables optical nonlinearities that extend all the way down to the single-photon level [15-19].

Radiative properties of two-dimensional (2D) excitons—bound pairs of an electron and a hole — can be drastically modified by placing them above a mirror [20,21]. Light emitted in the upward direction interferes with the light emitted downward and subsequently reflected by the mirror. Depending on the phase of the reflected light relative to the directly emitted light, the interference can be constructive, which leads to enhanced emission and a shortened radiative lifetime of the excitons, or destructive, which prolongs the radiative lifetime. While similar effects have been observed for individual dipolar optical emitters, such as atoms [22-24], ions [25], molecules [26-28] and quantum dots [29], near a mirror, the resultant changes in radiative rates has been limited to a few tens of percent because only a small fraction of optical modes is reflected by the mirror onto the quantum emitter. In contrast, the in-plane momentum of delocalized excitons in a homogeneous, atomically thin membrane is conserved during radiative decay such that light reflected by a planar mirror interacts deterministically with the same excitonic mode. As such, it is possible to achieve enhanced control over the radiative properties of 2D excitons using a mirror, including near complete suppression of radiative decay when the reflection loss is small [20,30].

The experimental realization of this conceptually simple phenomenon imposes stringent requirements on the excitons. First, excitons should exhibit excellent spectral and spatial uniformity and coherence, since any disorder and nonradiative losses suppress the interference [21]. Moreover, the excitons must be confined to a layer that is much thinner than their wavelength, and their in-plane momentum needs to be conserved throughout the process. Transition metal dichalcogenide (TMD) monolayers, a class of atomically thin direct-bandgap semiconductors, provide a platform for realizing the exciton-over-mirror system that meets many of these requirements. In particular, TMD monolayers encapsulated between two insulating hexagonal boron nitride (hBN) layers host delocalized excitons whose radiative decay rate can be much larger than nonradiative decay or pure dephasing rates [21,31-33]. Finally, when these heterostructures are suspended over a substrate, the spatial inhomogeneity that is often present in on-substrate heterostructures [32,34] can be alleviated, thus removing an important roadblock for preparing excitons coherent over micrometer length scales (see below).

Figure 1 illustrates the effect of hBN encapsulation and suspension on improving the spectral properties of $MoSe_2$ monolayers. In the device **D1** (Figs. 1(a) and 1(b)), we first prepare a hBN/$MoSe_2$/hBN heterostructure and then add graphene layers (less than 5 layers thick) to the bottom or top (or both sides) for electrostatic gating [21,35,36]. As shown previously [21,31-33], photoluminescence (PL) spectra of the heterostructure under off-resonant laser excitation (1.88 eV) at $T$ = 4 K exhibit exciton features with narrow linewidths both in the suspended and non-suspended regions (Fig. 1(c)). Importantly, whereas the PL peak positions from the regions on top of the $SiO_2$ substrate vary significantly from location to location (dashed lines in Fig. 1(c)), the regions suspended over a trench exhibit spatially homogeneous spectra in which the PL peaks overlap within the linewidth (~1.5 meV in **D1**) over the entire region (solid lines in Fig. 1(c)). We note that the suspended heterostructures exhibit improved spatial homogeneity whether there are graphene gates or not (Fig. S1 [37]). This observation suggests that the suspension reduces spatial inhomogeneity both by decoupling excitons from surface charge puddles [32,34] and by relieving the localized strain variations within the heterostructure [38].

The suspended TMD heterostructures with narrow, spatially homogeneous exciton transitions enable in-depth interrogation and control of the exciton properties. First we measure the absorption and PL spectra of the heterostructure as a function of gate voltage ($V_g$) applied between the graphene and the MoSe$_2$ monolayer (Figs. 1(d) and S1 [37]). As shown in Fig. 1(d), when the MoSe$_2$ monolayer is intrinsic ($V_g < 0V$ in **D1**), the reflection spectrum of the suspended heterostructure reveals sharp transitions that can be assigned to the excited Rydberg states of the A and B excitons. By fitting the observed energies of the Rydberg states to the screened Keldysh potential [39,40], we determine the bandgap and exciton binding energies of hBN-encapsulated MoSe$_2$ monolayers: specifically, the binding energies of the A and B 1s excitons are 0.192 eV and 0.210 eV, respectively. These values suggest a larger reduced mass of the B excitons, consistent with DFT calculations [41].

Next, we realize a graphene/hBN/MoSe$_2$/hBN/graphene heterostructure suspended above a metallic substrate by first depositing a gold layer inside the trench before transferring a heterostructure over it. The metal layer functions both as a mirror and as a local electromechanical gate. We apply a voltage ($V_B$) between the mirror and the bottom graphene layer to control their separation ($z$), as shown in Fig. 2(a). The application of $V_B$ produces an attractive force that reduces $z$. Importantly, in this doubly gated structure we can apply a compensating gate voltage between the top graphene and the MoSe$_2$ monolayer to avoid the MoSe$_2$ monolayer from being electrostatically doped.

Figures 2(b) and 2(c) display spatial images of spectrally integrated PL for an encapsulated MoSe$_2$ device, **D2**, measured at two different $V_B$. In this device, the trench depth is slightly larger than half the wavelength of the excitonic transition. The PL intensity from the region suspended above gold (enclosed by the dashed line) decreases dramatically as we change $V_B$ from 0 V to -150 V, whereas the intensity from the on-substrate region remains the same. We note that we observe the same effect in the PL intensity when we apply positive $V_B$, as expected for the electromechanical effect, and we therefore only discuss the system's response under negative $V_B$ in the following sections. Figures 2(d) and 2(e) show the PL spectra as a function of $V_B$, collected from the center of the suspended region: the integrated PL intensity decreases by a factor of ~10 with increasing $|V_B|$ despite the fact that the overall absorption of the device at 660-nm excitation is reduced by only ~10% [37]. Remarkably, the total PL linewidth of the exciton peak decreases from 2.2 meV to 1.3 meV in the same $V_B$ range while the exciton energy redshifts by ~4 meV (Figs. 2(d), 2(e) and S3 [37]).

The absolute reflectance spectra of **D2** at the center of the suspended region exhibit a similar trend (Figs. 3(a) and 3(b)). At $V_B$ = 0V, the reflectance spectrum exhibits a strong dip due to the neutral exciton absorption. Applying a larger magnitude of $|V_B|$ results in diminished absorption as well as much narrower linewidth. We note that all the reflectance spectra in Fig. 3(a) feature only one dip associated with the neutral exciton X$_0$, signifying that the MoSe$_2$ monolayer remains intrinsic over the entire $V_B$ range [35,42,43]. This observation indicates that the changes in reflectance and PL seen in Figs. 2 and 3 are not due to doping but rather originate from the change in the radiative properties of the neutral excitons X$_0$.

The $V_B$-dependent modification of the reflectance and PL properties of suspended heterostructures can be qualitatively understood in terms of the aforementioned interference effect between the field directly emitted by excitons and the field that is reflected by the mirror. By changing the distance $z$ between the TMD and the mirror with the applied voltage $V_B$, we vary the relative phase between the two components, thus modulating the radiative emission rate of excitons. The interference is destructive (constructive) and the radiative lifetime prolonged (shortened) if the distance between the TMD and the mirror is close to an even (odd) integer multiple of $\lambda/4$, where $\lambda$ is the wavelength of the exciton transition (the exact distances depend on the skin depth of the metallic mirror and the dielectric environment [37]).

Additional control experiments further support this interference picture. First, the $V_B$-dependent PL spectrum from the suspended region without the bottom mirror shows only an exciton energy shift but not the change in the PL intensity nor the linewidth (Fig. S5 [37]). These observations suggest that whereas the exciton energy shift originates from the strain-induced reduction of the MoSe$_2$ bandgap [44] (Figs. S3 & S4 [37]), the linewidth narrowing is mainly due to the modification of the local electromagnetic environment induced by the mirror. Second, when the initial separation of the monolayer and the bottom mirror is between $\lambda/4$ and $\lambda/2$, we observe an increase of the total exciton linewidth as the distance decreases (Fig. S6 [37]), consistent with the physical picture.

To gain a more quantitative insight, we model the total reflectance of our device using a master equation describing the exciton dynamics [21,37]. The reflection coefficient of any stack can be written as a function of the exciton energy $\hbar\omega_1$, total linewidth $\hbar\gamma$ and the background reflection $r_0$, which is the reflection from everything except the TMD monolayer (see Supplemental Material [37]). Based on Eq. (S8), we can extract these three parameters from each spectrum at a given $V_B$ (Fig. 3(c)). As the magnitude of the voltage increases, the extracted exciton total linewidth decreases from ~2.4 meV to 1.0 meV (Fig. 3(c)), consistent with PL measurements. Next, we compute the distance $z$ from $r_0$ using a transfer matrix approach as well as the known thicknesses and dielectric constants of the individual layers (Fig. 3(c), inset). The estimated $z(V_B)$ exhibits excellent agreement with a finite element simulation of the electromechanical response of the system, reaching a maximum deflection of the heterostructure of ~21 nm at the center of the trench at $V_B$ = -100 V. Below $V_B$ = -110 V, the measured background reflection exceeds the maximum reflection expected from the model, and therefore we restrict our analysis to voltages above V= -110 V. This discrepancy below $V_B$ = -110 V may arise from uncertainty in the exact values of the materials' dielectric functions (*e.g.* gate dependent refractive index of graphene).

The $z$-dependence of linewidth in Fig 3 (c) can be quantitatively explained by a change in the radiative linewidth of the excitons. As shown in Supplemental Material [37], the total linewidth may be written as $\hbar\gamma = \hbar\gamma_{r,vac}$ Re $g_0 + \hbar\gamma_{nr}$, where $\gamma_{r,vac}$ is the radiative decay rate in vacuum, $g_0$ is a function of the device geometry, and $\hbar\gamma_{nr}$ is the non-radiative contribution to the linewidth such as non-radiative decay and pure dephasing [21]. By computing the value of $g_0$ as a function of $z$ for a given device, we can obtain an excellent fit of the total linewidth (Fig. 3(d) and the values for $\gamma_{r,vac}$ and $\gamma_{nr}$. The fitted value of $\hbar\gamma_{r,vac}$= 1.33 meV (red line in Fig. 3(d)) agrees well with previous reports [21,45-47]. Figure 3(d) further shows the radiative linewidth $\hbar\gamma_r = \hbar\gamma_{r,vac}$ Re $g_0$ renormalized by the environment and the non-radiative broadening $\hbar\gamma_{nr} = \hbar(\gamma - \gamma_r)$. The renormalized radiative rate $\gamma_r$ varies by almost one order of magnitude from ~1.6 meV to 0.19 meV, as the distance $z$ decreases, while the non-radiative rate $\hbar\gamma_{nr}$ stays at 0.7 meV. This reduction in radiative rate is consistent with the decrease in PL intensity.

We note that factors not considered by the model, such as curving of the heterostructure membrane and gate-dependent refractive index of graphene, prevent us from reliably distinguishing the pure dephasing from nonradiative rate [21]. For instance, the variation in $z$ across the trench due to the curving of the heterostructure can result in spatially varying radiative lifetime of excitons and can in turn have a significant impact on the observed lineshape, particularly under large $|V_B|$ (we estimate the variation of $z$ to be ~4 nm within a 1 μm laser spot at the center of the heterostructure at $V_B$ = -100 V for **D2** (Fig. S4 [37])). This may lead to systematic errors when extracting the pure dephasing rate by fitting the exact reflection lineshape for large $|V_B|$. In the above analysis we mitigate these effects by focusing on more robust metrics such as the total linewidth and the radiative rate.

Because our suspended heterostructure exhibits an intrinsic mechanical resonance, we can perform optical measurements while driving the membrane's mechanical motion with an AC voltage. First, we characterize the motion of the heterostructure via interferometry using an off-resonant continuous-wave laser while varying the frequency of the AC drive to the metal gate, $V_B$. The device **D3** exhibits a fundamental mechanical resonance frequency $f$ = 17.97 MHz with a quality factor of ~650 under a 0.5 V peak-to-peak AC voltage (inset, Fig. 4(a)). The quality factor decreases with increasing AC amplitude, indicating nonlinearity of the electromechanical response in this compound system (Fig. S7 [37]).

The maximum deflection of the heterostructure under resonant AC drive can be much larger than that under DC drive with the same voltage amplitude. Thus, we can utilize the mechanical resonance to induce a more significant change in the radiative properties of excitons. Figure 4(a) shows the time-resolved PL emission intensity from **D3** under various driving frequencies with a 10 V peak-to-peak AC voltage. When the driving frequency is resonant with the fundamental mechanical mode of the suspended heterostructure, we realize more than 50% modulation of the PL intensity. In contrast, with a DC voltage of equivalent amplitude (5 V), the modification of PL intensity in **D3** is less than 5%. Similarly, the reflectance spectrum from the heterostructure exhibits a significant temporal modulation under resonant AC drive: the absolute reflectance at the exciton resonance changes by more than 70% (Fig. 4(b)). From the time-dependent reflectance spectrum, we estimate that the maximum deflection of the membrane is 14 nm, and the radiative linewidth is modulated by ~60% during the mechanical oscillation.

Suspended heterostructures based on atomically thin semiconductors feature mechanical, electrical, and optical degrees of freedom that are intertwined, and open up a plethora of new applications. Owing to the excellent optical coherence demonstrated in this work, as well as the mechanical degrees of freedom, suspended heterostructures represent an attractive platform for exploring the interaction between electromagnetic radiation and nanomechanical motion [48]. Importantly, the system naturally exhibits both dispersive and dissipative optomechanical coupling, whose relative strength can be tuned by the distance between the heterostructure and the mirror. Dissipative coupling may enable ground-state cooling and quantum limited-position detection even when the mechanical sideband is not resolved [49,50]. The TMD-based mechanical resonators with high resonance frequencies and high quality factors [51,52] should provide new opportunities in this regard, as well.

The demonstrated reduction in the total linewidth (down to ~0.8 meV in Fig. S8 [37]), may already find new applications such as in topological photonics, which require a narrow exciton linewidth to split the exciton by over a linewidth with a magnetic field [53]. Manipulating the radiative processes of the emitters at a timescale faster than their lifetime can facilitate new methods to store and release photons on-demand. This may be achieved by using, for example, interlayer excitons with longer lifetimes [54,55] and material systems hosting interlayer excitons with large enough oscillator strength [56]. The lifetime of the excitons can be extended by a factor of $1/\rho$, where $\rho$ is the reflection loss of the metallic mirror, leading to significantly enhanced optical nonlinearity [20]. Moreover, the high coherence of excitons [21] and strong interactions between Rydberg excitons [57] in heterostructures open up exciting new avenues for optoelectronics, active metasurfaces [58], topological photonics [53] and quantum nonlinear optics [20,59].

ACKNOWLEDGMENTS

We acknowledge support from the DoD Vannevar Bush Faculty Fellowship (N00014-16-1-2825 for HP, N00014-18-1-2877 for PK), NSF (PHY-1506284 for HP and MDL), NSF CUA (PHY-1125846 for HP and MDL), AFOSR MURI (FA9550-17-1-0002), ARL (W911NF1520067 for HP and MDL), the Gordon and Betty Moore Foundation (GBMF4543 for PK), ONR MURI (N00014-15-1-2761 for PK), and Samsung Electronics (for PK and HP). The device fabrication was carried out at the Harvard Center for Nanoscale Systems.

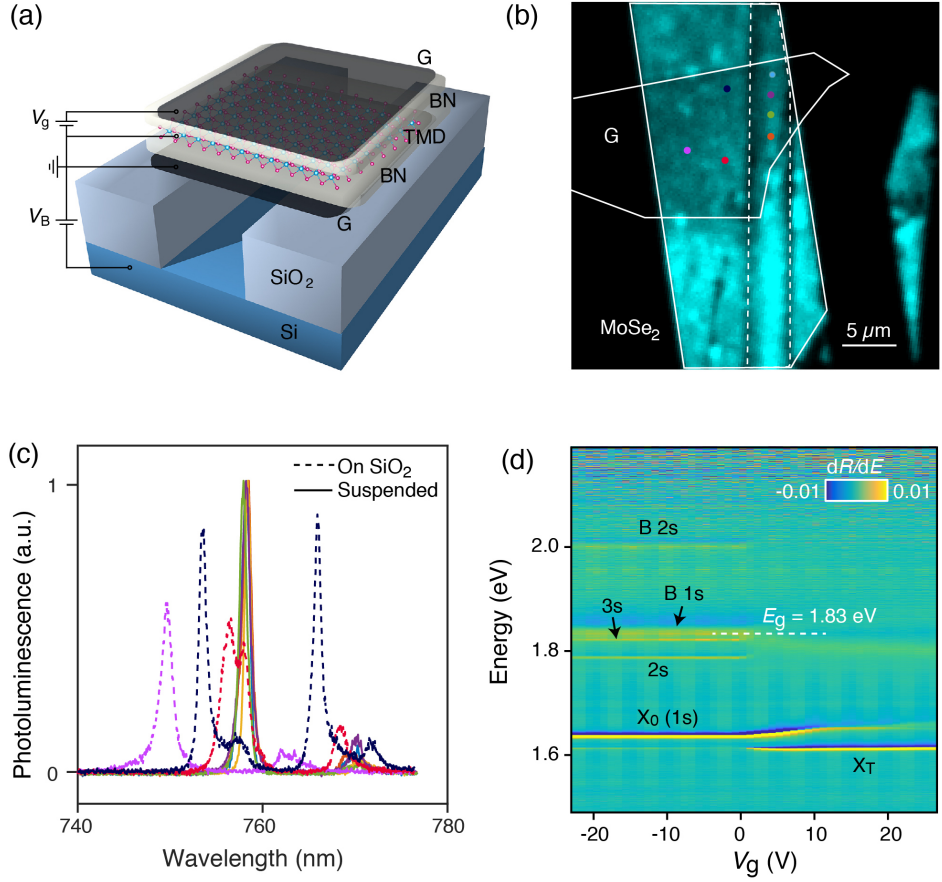

**FIG. 1**. **Spectroscopic characterization of suspended TMD-based van der Waals heterostructures at $T$ = 4 K.** (a) Schematic of a suspended heterostructure. The doping concentration of the TMD is electrostatically controlled through graphene gates by $V_g$. (b) A spatial PL map of a graphene/hBN/MoSe$_2$/hBN heterostructure device **D1**, where the suspended region is enclosed by the dashed line, and the enclosed area labeled G is where graphene is present. (c) Representative PL spectra taken from specific spots in (b). The solid lines correspond to the suspended region, while dashed lines are collected from spots on the substrate (color coded in the same way as in (b)). The suspended region shows improved spatial homogeneity with an exciton energy variation of less than the total linewidth. (d) Gate dependent differential reflectance spectrum of a suspended graphene/hBN/MoSe$_2$/hBN device, **D1**. The MoSe$_2$ monolayer is intrinsic at $V_g$ < 0V and becomes *n*-doped under positive $V_g$. The spectrum in the intrinsic regime shows clear absorption features from A and B neutral excitons as well as their excited states [36,40]. The band gap (white dotted line) is calculated based on the binding energy obtained from the Keldysh potential [36,40].

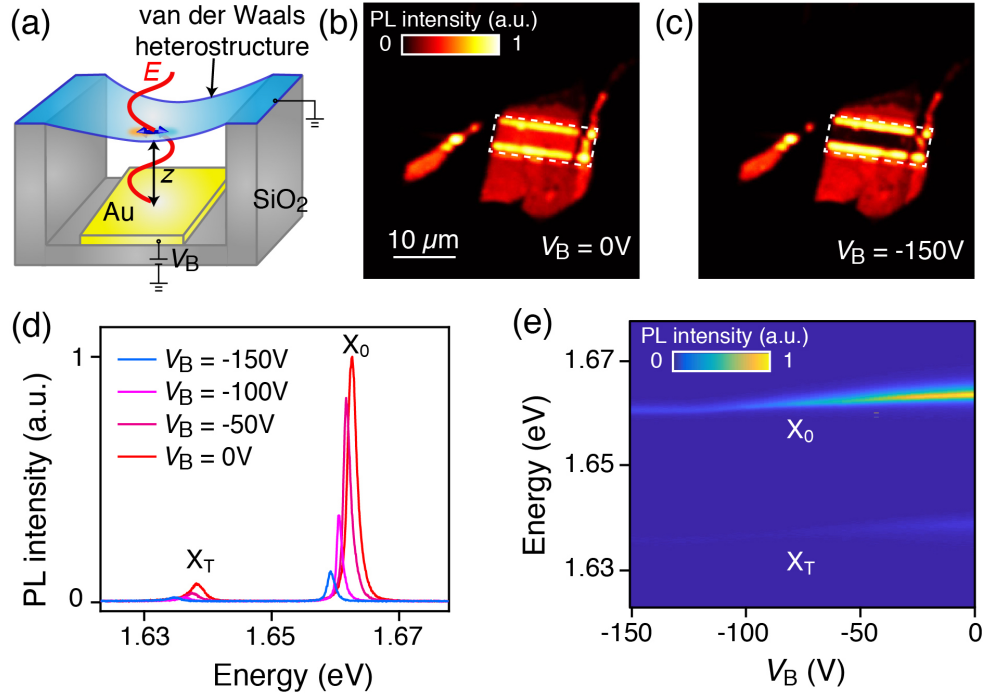

FIG. 2. **Electromechanical control of exciton emission for device D2 incorporating a suspended graphene/hBN/MoSe$_2$/hBN/graphene heterostructure.** (a) A schematic of the heterostructure suspended over a gold mirror. A voltage, $V_B$, is applied to vary $z$, the distance between the heterostructure and the mirror, which modifies the local electromagnetic environment of the excitons. (b,c) Integrated PL maps of **D2** under different $V_B$. A clear modulation of PL intensity can be observed in the suspended region (enclosed by the dashed lines) above the mirror but not in the rest of the sample. There is no bottom gold mirror at the edge of the suspended region as shown in (a), which leads to enhanced PL radiation intensity compared with the central region where there is a bottom gold mirror. (d) 1D cuts and (e) 2D plot of PL spectrum as a function of $V_B$ taken from a spot in **D2** showing a modulation of the PL intensity, linewidth, and energy of the excitons.

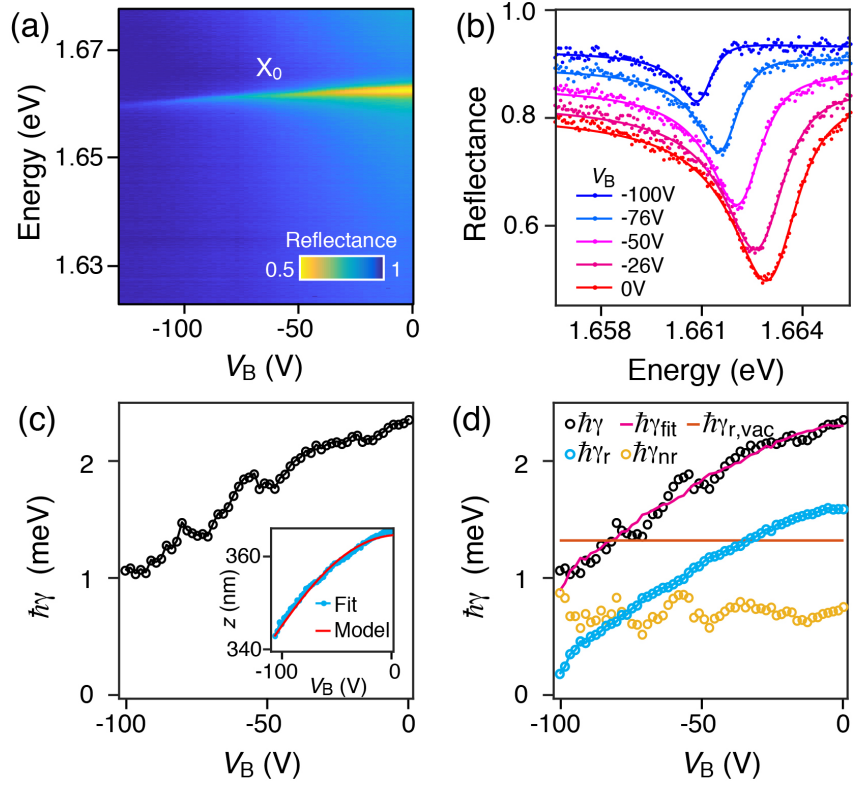

**FIG. 3. Electromechanical manipulation of excitonic reflectance at $T$ = 4 K for D2 incorporating a suspended graphene/hBN/MoSe$_2$/hBN/graphene heterostructure.** (a) 2D map and (b) linecuts of reflectance spectra of **D2** under various $V_B$ showing a modulation of the linewidth as well as the energy of the excitons. The data are shown as dots with fits overlaid as solid lines in (b). (c) The values of the total linewidth $\hbar\gamma$ (black circles) extracted from the fit of the reflection spectra show a large change as a function of $V_B$. Inset: the fitted height $z$ from reflectance measurements as a function of $V_B$ shows excellent agreement with a finite element electromechanical model. (d) By further modeling the gate-dependence of $\hbar\gamma$ (the fit shown as solid magenta line), we can extract the vacuum radiative linewidth in $\hbar\gamma_{r,\text{vac}}$ (solid horizontal red line). Together with the extracted height $z$, the value of $\hbar\gamma_{r,\text{vac}}$ can be used to calculate the radiative linewidth $\hbar\gamma_r$ (blue circles) and the nonradiative linewidth $\hbar\gamma_{nr}$ (yellow circles).

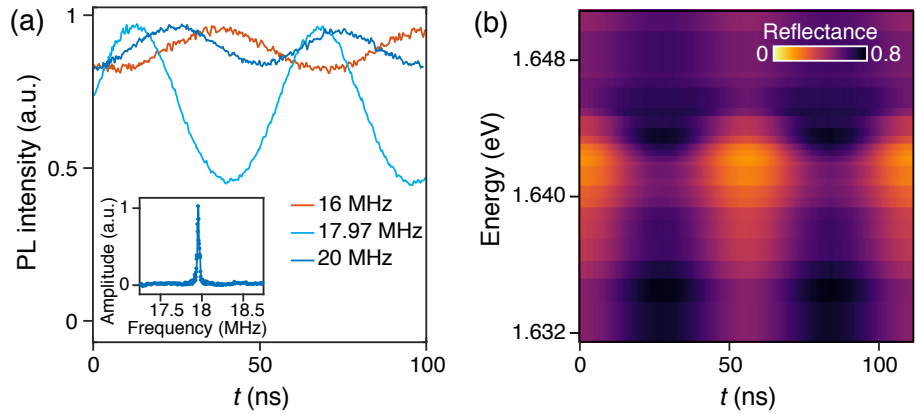

**FIG. 4. Modulation of exciton emission and reflectance of device D3 incorporating a suspended graphene/hBN/MoSe$_2$/hBN/graphene heterostructure, at $T$ = 4 K under AC electrical drive.** (a) Dynamic modulation of PL emission under various driving frequencies. Inset: amplitude of mechanical oscillation as a function of electrical drive frequency showing the mechanical resonance of **D3**. (b) Dynamic modulation of reflection near exciton resonance with a resonant AC drive.